\documentclass[groupedaddress,
  aps,
  pra,
  amsmath,amssymb,
  twocolumn,reprint
]{revtex4-1}
\usepackage{graphicx,color}
\usepackage{amsmath}
\usepackage{natbib}
\usepackage{epsfig}
\begin{document}

\title{{\color{blue} Theory of a cavity around a large floating sphere in complex (dusty) plasma }}

\author{Sergey Khrapak}
\email{Sergey.Khrapak@dlr.de}
\affiliation{Institut f\"ur Materialphysik im Weltraum, Deutsches Zentrum f\"ur Luft- und Raumfahrt (DLR), 82234 We{\ss}ling, Germany}

\author{Peter Huber}
\affiliation{Institut f\"ur Materialphysik im Weltraum, Deutsches Zentrum f\"ur Luft- und Raumfahrt (DLR), 82234 We{\ss}ling, Germany}

\author{Hubertus Thomas}
\affiliation{Institut f\"ur Materialphysik im Weltraum, Deutsches Zentrum f\"ur Luft- und Raumfahrt (DLR), 82234 We{\ss}ling, Germany}

\author{Vadim Naumkin} 
\affiliation{Joint Institute for High Temperatures, Russian Academy of Sciences, 125412 Moscow, Russia}

\author{Vladimir Molotkov}
\affiliation{Joint Institute for High Temperatures, Russian Academy of Sciences, 125412 Moscow, Russia}

\author{Andrey Lipaev}
\affiliation{Joint Institute for High Temperatures, Russian Academy of Sciences, 125412 Moscow, Russia}

\date{\today}

\begin{abstract}
In the last experiment with the PK-3 Plus laboratory onboard the International Space Station, interactions of millimeter-size metallic spheres with a complex plasma were studied~[M. Schwabe {\it et al.}, New J. Phys. {\bf 19}, 103019 (2017)]. Among the phenomena observed was the formation of cavities (regions free of microparticles forming a complex plasma) surrounding the spheres. The size of the cavity is governed by the balance of forces experienced by the microparticles at the cavity edge. In this article we develop a detailed theoretical model describing the cavity size and demonstrate that it agrees well with sizes measured experimentally. The model is based on a simple practical expression for the ion drag force, which is constructed to take into account simultaneously the effects of non-linear ion-particle coupling and ion-neutral collisions. The developed model can be useful for describing interactions between a massive body and surrounding complex plasma in a rather wide parameter regime.    
\end{abstract}

\maketitle

\section{Introduction}

Understanding fundamental interactions between an object and surrounding plasma is an exceptionally important problem with application to astrophysical
topics~\cite{Goertz1989,ShuklaBook2002}, plasma technology~\cite{BouchouleBook1999}, plasma medicine~\cite{KOngNJP2009}, complex (dusty) plasmas~\cite{FortovUFN2004,FortovPR2005,MorfillRMP2009} and fusion related problems~\cite{WinterPPCF1998}.  Considerable progress on the interaction of micron-size plastic particles with weakly ionized plasma medium has been achieved thanks to complex plasma research program under microgravity conditions onboard the International Space Station (ISS). This particularly concerns particle charging, the ion drag force, interparticle interactions, linear and non-linear wave phenomena, see for instance Ref.~\cite{ThomasPPCF2019} for a recent review.   

Here another related problem is addressed, namely how a bigger object interacts with surrounding complex plasma. New information about these interactions has been obtained from the last experimental campaign with PK-3 Plus laboratory onboard ISS~\cite{SchwabeNJP2017}. In these experiments the metallic spheres of one millimeter diameter were injected into a low-temperature rf discharge together with microparticles forming a complex plasma. Various phenomena were observed, including motion of spheres through a complex plasma cloud, generation of bubbles, ``repulsive attraction'', and excitation of low-frequency waves~\cite{SchwabeNJP2017}. 

It was also observed that when a sphere passes through a complex plasma cloud, it is surrounded by a cavity of a few millimeter in diameter, where no microparticles are present. It is the size of the cavity which is the main object of interest here. The size of the cavity is relatively easy to measure and it contains important information about the system parameters. The size obviously depends on the balance of forces acting on the particles located at the cavity edge. The main forces identified are the short-range electric repulsion from the highly charged sphere and the long-range attraction triggered by the ion flow (the ion drag force), which is directed towards the sphere surface~\cite{SchwabeNJP2017}.  In this article we first propose a simple practical expression for the ion drag force for the conditions relevant for the experiment. In particular, this expression allows us to take into account simultaneously the effects of non-linear ion-particle coupling and ion-neutral collisions. Then, using this expression, we formulate the force balance condition and estimate theoretically the cavity diameter. We show that the estimated diameter agrees well with the results of experimental measurements.

The theoretical approximation developed here should be applicable (possibly with some modifications) to other situations corresponding to the interaction of large objects with complex plasmas, such as for instance probe-induced voids and particle circulations~\cite{LawPRL1998,ThompsonPoP1999,SamsonovPRE2001,ThomasPoP2004,KlindworthPRL2004,HarrisPRE2015}, as well as the formation of boundary-free clusters~\cite{UsachevPRL2009}.   

\section{Experiment}

The experiment to be discussed is the last experiment of the PK-3 Plus laboratory, which operated onboard the ISS in 2006-2013~\cite{ThomasPPCF2019,ThomasNJP2008,KhrapakCPP2016}. The experiment is described in detail in Ref.~\cite{SchwabeNJP2017}. Here we provide only the brief summary, necessary for the understanding of this article.

\begin{figure}
\includegraphics[width=8cm]{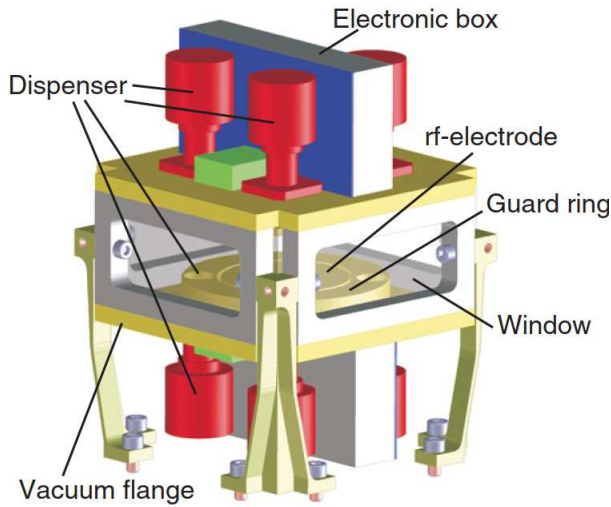}
\caption{Sketch of the PK-3 Plus discharge chamber. Adapted from ~\cite{ThomasNJP2008}.}
\label{Fig1}
\end{figure}  

The PK-3 Plus laboratory consisted of a radio-frequency plasma chamber with two electrodes of $6$ cm in diameter separated by a distance of 3 cm. The electrodes were surrounded by grounded guard rings, see sketch in Fig. 1. Dispensers mounted in the guard rings were used to introduce microparticles of
various sizes into the gas discharge. Highly charged microparticles formed large symmetric three-dimensional clouds in the plasma bulk. Typically, these clouds contained a central particle-free region -- the so-called ``void'' -- attributed to the action of the ion drag force, pushing the particles to the periphery~\cite{ThomasPPCF2019,GoreePRE1999,GozadinosNJP2003,KretschmerPRE2005,LipaevPRL2007,GoedheerPPCF2008}.  Strong interparticle interactions between microparticles resulted in structures typical for the fluid and solid states. Structural and dynamical properties of the particle component were studied at the most fundamental kinetic level, providing new insight into the physics of a new plasma state of soft condensed matter~\cite{ChaudhuriSM2011}.     

\begin{figure}
\includegraphics[width=8cm]{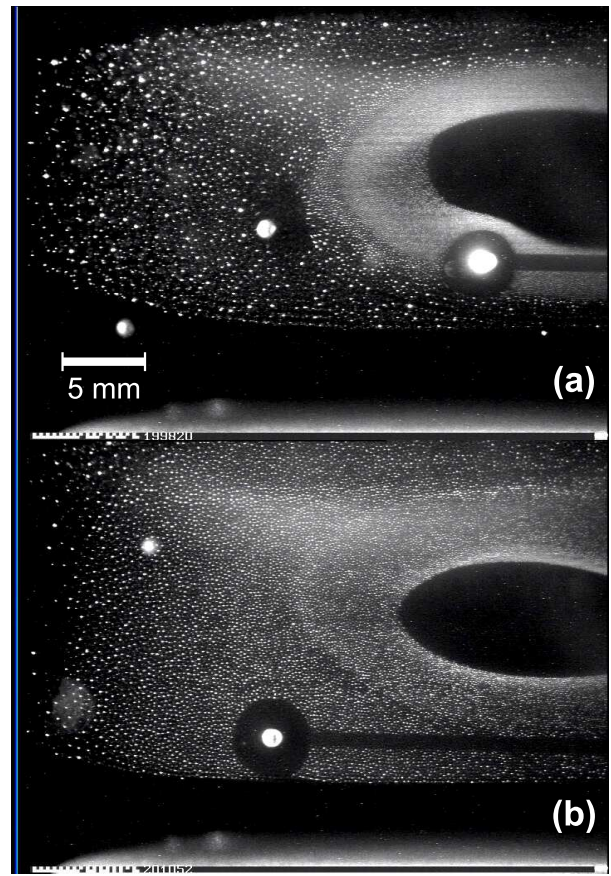}
\caption{Experimental video-images showing a metallic sphere surrounded by complex plasma in an argon discharge.  In (a) the pressure is 17.5 Pa, the particles interacting with the sphere have a diameter of $1.55$ $\mu$m, the diameter of the cavity is $\simeq 4.2$ mm; in (b) the pressure is 30.4 Pa, complex plasma interacting with the sphere consists mostly of agglomerate particles, the diameter of the cavity is $\simeq 4.8$ mm. }
\label{Fig2}
\end{figure}

Many important fundamental phenomena were studied using PK-3 Plus laboratory, including, for instance, equilibrium and non-equilibrium phase transitions~\cite{WysockiPRL2010,SutterlinPPCF2010,KhrapakPRL2011,KhrapakPRE2012}, lane formation~\cite{DuNJP2012}, wave excitation~\cite{SchwabeNJP2008}, instabilities~\cite{SchwabePRE2017,HeidemannPoP2011}, Mach cones~\cite{SchwabeEPL2011,ZhukhovitskiiPoP2015}, etc.

In the experiment discussed here interactions between millimeter size metallic spheres and complex plasmas under microgravity conditions were studied. Previously, penetration of complex plasma clouds by fast charged projectiles was already investigated under microgravity conditions~\cite{ArpPRE2011,CaliebePoP2011}. In particular, the dynamics of the formation of an elongated cavity in the projectile's wake was analyzed in detail. Present work deals with much larger and slower objects interacting with complex plasma.  

For the purpose of the experiment, the dispensers were shaken so strong that the metallic spheres of 1 mm diameter that were present inside the dispensers broke through the sieve and entered the bulk plasma region together with the microparticles remaining in the dispensers~\cite{SchwabeNJP2017}. Furthemore, the cosmonaut Pavel Vinogradov, who performed the experiment, shook the experimental container to impact momentum on the spheres. The shaking had little effect on the plasma and microparticles, but accelerated spheres by collisions with chamber walls. As a result, the spheres experienced an almost force-free motion inside the discharge chamber~\cite{SchwabeNJP2017}.  

The analysis of the motion of spheres through complex plasma clouds was reported previously~\cite{SchwabeNJP2017}. Here the main interest is to the size of the cavities that are created around the spheres. The size can be relatively easily estimated for events when the spheres cross the plane formed by the laser sheet used to illuminate complex plasma.
Two such exemplary events are shown in Fig.~\ref{Fig2}. About twenty such events have been analyzed and the sizes of the cavities have been estimated as follows. The crossing of a laser sheet by the sphere corresponds to several video frames. From these frames a single frame with the largest cavity size is selected. After correcting the aspect ratio of the video frame, it has been verified that the observed cross sections of the cavities have a shape close to a circle. In a graphical editor, a circle has been selected that fits most accurately into the cavity observed on the frame. The diameter of this circle is used as an estimate of the cavity diameter. The relative error in estimating the diameter does not exceed 5 $\%$. More accurate evaluation of the cavity size and shape can be made from a careful analysis of three-dimensional trajectories of the spheres, as described in Ref.~\onlinecite{SchwabeNJP2017} for a single crossing event, but this is not necessary for the present purpose.   

The sizes of the cavities have been measured for different neutral gas pressures  (in the range between $\simeq 15$ and $\simeq 30$ Pa) and for situations where the spheres were interacting with microparticles of different sizes (complex plasma cloud consisted of particles with diameters of 1.55, 2.55, 3.42, 6.8, 9.19, and 14.9 $\mu$m as well as their agglomerates; this mixture was heterogeneous with smaller particles located closer to the discharge center and bigger particles pushed further to the periphery, see Fig~\ref{Fig2}). It has been observed that the cavity size increases with pressure, but is practically insensitive to the microparticles size with which sphere is interacting. These trends correlate well with the results of theoretical consideration performed below.                
 
\section{Theory}    

When a sphere is immersed into a plasma it starts to collect electrons and ions on its surface, just as smaller microparticles do. As a result both spheres and microparticles are charged negatively, the surface (floating) potential being roughly of the order of the electron temperature, which ensures that the ion and electron fluxes to the surface can balance each other. If the sphere is surrounded by a complex plasma, the particles experience the following forces. At short distances there is a strong electrostatic repulsion of negatively charged particles from the negatively charged sphere. At sufficiently long distances from the sphere, the ion drag force associated with the ion flux towards the sphere surface can overcome the electrostatic repulsion. There can be also the pressure force exerted by the microparticle cloud, directed towards the metallic sphere. This, however, was shown to be numerically small for typical experimental conditions~\cite{SchwabeNJP2017} and will not be considered.     

Our main assumption is that the cavity boundary position is mainly determined by the balance between the electric repulsion at short distances and the ion-drag-mediated attraction at long distances. In the following we consider the force balance for an individual microparticle located at an equilibrium position, where both forces compensate each other. In this way we neglect (i) the effect of particles on the distribution of the electrostatic potential around the sphere and (ii) some reduction of the ion drag force in dense dust clouds~\cite{YaroshenkoPoP2013}. Both assumptions are reasonable in not too dense microparticle clouds as those observed in the experiment.

The ratio of the ion drag to the electric forces, $F_{\rm i}/F_{\rm el}$, is known to be approximately constant for subthermal ion flows and then to decrease relatively fast in the super-thermal regime~\cite{KhrapakPoP2005}. For this reason, the cavity boundary should be roughly located at a position where $M=u/v_{{\rm T}i}\sim 1$, where $u$ is the ion drift velocity, $v_{{\rm T}i}$ is the ion thermal velocity, and $M$ is the ion thermal Mach number. This implies that the perturbations created by a large floating metallic sphere at the position of the boundary are relatively small (much smaller than in the sheath region formed around the sphere surface, where the ion drift is super-sonic). This suggests to focus on the long-range asymptote of the electric potential generated by a large floating body and not on the plasma properties in its immediate vicinity. In this comparatively far region the effects associated with plasma absorption on the sphere govern the distribution of the electric potential and this simplifies considerably the consideration, as we will see below. The first step, however, is to develop an appropriate model for the ion drag force.     

\subsection{Ion drag force}

In the parameter regime investigated, the characteristic length scale of ion-particle interactions exceeds the plasma screening length, indicating that ion-particle interactions are non-linear. Several theoretical approaches have been developed for this regime, mostly using binary collision approximation~\cite{KhrapakPRE2002,KhrapakPRL2003,KhrapakIEEE2004,KhrapakPoP2014,SemenovPoP2017}. However, these purely collisionless treatments are not very appropriate for our purpose, because ion-neutral collisions can be important in the pressure range investigated~\cite{KhrapakCPP2009}. Collisional effects can be incorporated into kinetic or hydrodynamic calculations using the linear plasma response formalism~\cite{IvlevPRL2004,IvlevPPCF2004,IvlevPRE2005,KhrapakJAP2007}. Unfortunately, as we have just discussed, the linear approximation is not justified in present conditions (as well as in most other complex plasma experiments), because of significant non-linearities in ion-particle interactions. An approach, which accounts for both non-linearity in ion-particle interactions and the effect of ion-neutral collisions is required. 

Recently, the ion drag force has been calculated self-consistently and non-linearly using particle in cell codes, taking into account ion-neutral collisions~\cite{HutchinsonPoP2013}. These calculations demonstrated that the magnitude of the force is sensitive to the ion velocity distribution function for superthermal ion flows.  It was shown that the finite collisionality initially enhances the ion drag force up to a factor of 2 relative to the collisionless result. Larger collisionality eventually reduces the ion drag force, which can even reverse sign in the continuum limit~\cite{KhrapakJAP2007,FilippovJETPLett2008,VladimirovPRL2008,SemenovPoP2013,MomotPoP2017}, but this regime is too far from typical experimental conditions. Most important for our present purpose is that the collisional drag enhancement can be represented by an almost universal function of scaled collisionality and flow velocity, for which simple fits are available~\cite{HutchinsonPoP2013}.   

We pursue the following strategy. First, an {\it ad hoc} simple practical expression for the collisionless ion drag force, based on our earlier theoretical results from the binary collision approach, is derived. It is demonstrated to be in good agreement with the non-linear collisionless simulation results of Ref.~\cite{HutchinsonPoP2013}. Then a correction factor, expressing the influence of ion-neutral collisions on the ion drag force, as suggested in~\cite{HutchinsonPoP2013}, is added to the collisionless expression. This provides us with a new practical expression for the non-linear ion drag force in the collisional regime, which will be then used to estimate the size of the cavity around the metallic sphere.           

We start with an expression for the ion drag force derived for the regime of intermediate non-linearity~\cite{KhrapakPRE2002}  
\begin{equation}\label{id1}
F_{\rm i}=\left(8\sqrt{2 \pi}/3\right)a^2 n_i m_i v_{{\rm T}i}u\left(1+\frac{z\tau}{2}+\frac{z^2\tau^2}{4}\Lambda\right),
\end{equation}
where $\Lambda$ is the modified Coulomb logarithm
\begin{equation}\label{CoulLog}
\Lambda = 2\int_0^{\infty}e^{-x}\ln\left(\frac{2\lambda x/a+z\tau}{2x+z\tau}\right)dx.
\end{equation}
Other notation is as follows: $a$ is the particle radius, $n_i$, $m_i$, $T_i$, $v_{{\rm T}i}=\sqrt{T_i/m_i}$ are the ion density, mass, temperature, thermal velocity, $z=e|\phi_s|/T_e$ is the particle surface potential ($\phi_s$) expressed in units of the electron temperature $T_e$, $\tau=T_e/T_i$ is the electron-to-ion temperature ratio, and $\lambda$ is the effective plasma screening length. 

This expression applies to subthermal ion flows, $u\lesssim v_{{\rm T}i}$.  It can be considered a generalization of the standard Coulomb scattering theory, by taking into account the impact parameters beyond the plasma screening length: all ions which {\it approach} the grain closer than $\lambda$ are included in the consideration. Therefore, it is sometimes referred to as the modified Coulomb scattering approach. Quantitatively, the approach has been originally proposed for the regime $\beta = z\tau (a/\lambda)\lesssim 5$, where $\beta$ is known as the scattering parameter~\cite{KhrapakPRL2003}.  In the regime $\beta\ll 1$, it reduces to the conventional Coulomb scattering theory.

We can further simplify Eqs.~(\ref{id1}) and (\ref{CoulLog}) as follows.  We neglect the collection part of the momentum transfer [first two terms in brackets of Eq.~(\ref{id1})]. In the expression for the modified Coulomb logarithm we make use of the typical condition $z\tau\gg 1$ to arrive at
\begin{displaymath}
\Lambda \simeq 2\int_0^{\infty}e^{-x}\ln\left(1+2x/\beta\right)dx.
\end{displaymath}
Thus, the modified Coulomb logarithm depends mainly on $\beta$, and it is easy to demonstrate (by way of direct numerical integration)
that for $\beta\gtrsim 1$ the integral above can be very well approximated as $\Lambda\simeq 1.8\ln(1+2/\beta)$. In the non-linear regime considered this becomes simply $\Lambda\simeq 3.6/\beta$. This allows us to write
\begin{equation}\label{id2}
F_{\rm i}\simeq 6.0 a^2 n_i T_i M z\tau (\lambda/a).
\end{equation}   

This represents an expression for the non-linear ion drag force in the collisionless regime to be compared with numerical results from Ref.~\cite{HutchinsonPoP2013}. In that numerical investigation the particle surface potential as well as the electron-to-ion temperature ratio were fixed to $z=2$ and $\tau=100$, respectively. The ratio $\lambda_{{\rm D}e}/a$ varied in the range from 10 to 200, where $\lambda_{{\rm D}e}=\sqrt{T_e/4\pi e^2n_e}$ is the electron Debye radius. The ion drag force was expressed in units of $n_e T_e a^2$. To simplify the comparison we can rewrite Eq.~(\ref{id2}) as
\begin{equation}\label{id3}
(F_{\rm i}/n_eT_e a^2)\simeq 12.0 (\lambda_{{\rm D}e}/a)(u/c_{\rm s}),
\end{equation}   
where $c_s=\sqrt{T_e/m_i}$ is the ion sound velocity. In arriving to Eq.~(\ref{id3}) we assumed quasineutrality, $n_e=n_i=n_0$, and used the dominance of ion screening, $\lambda\simeq \lambda_{{\rm D}i}= \lambda_{{\rm D}e}/\sqrt{\tau}$. Note also that $u/c_{\rm s}=M/\sqrt{\tau}$. The obtained formula (\ref{id3}) demonstrates very close agreement with the numerical results presented in Figs. 8, 9 and 10(a) of Ref.~\cite{HutchinsonPoP2013}. Thus, the region of validity of the approximation (\ref{id2}), $\beta\sim {\mathcal O}(10)$, is somewhat expanded in the non-linear regime compared to the original approach (\ref{id1}) designed for $\beta\sim {\mathcal O}(1)$. Moreover, detailed comparison shows that it is reliable not only for the subthermal regime, but also for near-thermal and slightly superthermal ion flows (in the regime where difference in ion velocity distribution functions does not lead to considerable variations in the ion drag force). Further insight comes from the careful analysis of the data shown in Fig. 10(b) of ~\cite{HutchinsonPoP2013}, which demonstrates that in the collisionless limit Eq.~(\ref{id3}) remains accurate even at $M=2$ ($u/c_{\rm s} = 0.2$), provided the microparticles are not too small ( $\lambda_{{\rm D}e}/a\lesssim 50$).     

The collisional enhancement of the ion drag force can be expressed as a product of the collisionless force and a collisional correction factor~\cite{HutchinsonPoP2013} 
\begin{equation}\label{id4}
F_{\rm i} \simeq 6.0 a^2 n_i T_i M z\tau (\lambda/a) F({\tilde \nu}),
\end{equation}
with
\begin{equation}\label{factor}
F({\tilde \nu}) = \frac{1+A{ \tilde \nu}}{1+B{ \tilde \nu}+C{ \tilde \nu}^2},
\end{equation}
where ${ \tilde \nu} = \nu r_{\rm c}/c_{\rm s}$ is the reduced collisionality and $r_{\rm c}$ is the non-linear shielding cloud radius, derived in Ref.~\cite{HutchinsonPoP2013}. The latter is approximately 
\begin{displaymath}
r_{\rm c}\simeq 1.2\lambda_{{\rm D}e}\left(\frac{a}{\lambda_{{\rm D}e}}\frac{T_i}{Te}\right)^{1/5}.
\end{displaymath}
The coefficients provided in Ref.~\cite{HutchinsonPoP2013} for the  drift distribution of ion velocities (which is more appropriate for ions drifting through the stationary background of neutrals under the action of electric force, compared to a conventional shifted Maxwellian distribution) are
\begin{displaymath}
A=7+3M, \quad B=1.8 M, \quad C= 0.5 A.
\end{displaymath} 

Let us now compare the magnitudes of the electrostatic and ion drag forces in the limit of a weak electric field $E$, when the ion drift is subthermal.  The ion drift velocity is expressed 
\begin{equation}\label{drift}
u=\frac{eE}{m_i\nu_{\rm eff}},
\end{equation}
where $\nu_{\rm eff}$ is the effective collision frequency, which is field-dependent in general, but constant in the subthermal drift regime (weak electric field)~\cite{FrostPR1957,KhrapakJPP2013}. The electrostatic force is 
\begin{equation}\label{electric} 
F_{\rm el}= QE,
\end{equation} 
where $Q$ is the particle charge. The ratio of the collisionless ion drag force, Eq.~(\ref{id2}), to the electric force, Eq~(\ref{electric}), is then
\begin{equation}\label{ratio}
|F_{\rm i}/F_{\rm el}|\simeq 0.5 (\omega_{{\rm p}i}/\nu_{\rm eff})=0.5 (\ell_i/\lambda),
\end{equation}      
where $\omega_{{\rm p}i}=\sqrt{4\pi e^2 n_i/m_i}$ is the ion plasma frequency, $\ell_i=v_{{\rm T}i}/\nu_{\rm eff}$ is the ion mean free path with respect to collisions with neutrals. In deriving Eq.~(\ref{ratio}) it was assumed that screening is mostly associated with the ion component and, hence, $\lambda\simeq v_{{\rm T}i}/\omega_{{\rm p}i}$. For the particle charge we used $|Q|\simeq z (aT_e/e)$. Equation (\ref{ratio}) is very similar to that derived earlier in Ref.~\cite{KhrapakPRE2002}.
It can now be improved by taking ion-neutral collisions into account. An obvious modification reads
\begin{equation}\label{ratio1}
|F_{\rm i}/F_{\rm el}|\simeq 0.5 (\omega_{{\rm p}i}/\nu_{\rm eff})F(\tilde{\nu}).
\end{equation}      

The necessary condition of particle attraction to the sphere at long distances is $|F_{\rm i}/F_{\rm el}|>1$ in the limit of weak electric field. The electric field at which the ratio $|F_{\rm i}/F_{\rm el}|$ drops to unity will determine the cavity radius in this approximation.

Equations (\ref{id4}) and (\ref{factor}) represent an important intermediate result, providing new simple practical tool to evaluate the ion drag force under typical experimental conditions. We have a good opportunity to test it by comparing the predicted size of cavities with those observed experimentally.   

\subsection{Electric potential around sphere}

At sufficiently long distances from the sphere, the electric potential distribution is dominated by ion absorption on the sphere surface. The ion flux conservation allows to obtain the electric potential in the weakly perturbed quasi-neutral region. For a large sphere ($R_{\rm s}\gg \lambda$) and collision-dominated ion flux to its surface ($R_{\rm s}\gg\ell_i$) simple expressions for the potential and electric field are~\cite{RaizerBook} 
\begin{equation}\label{field}
\phi(r)\simeq -(T_e/e)(R_{\rm s}/r), \quad E(r)\simeq -(T_e/e)(R_{\rm s}/r^2),
\end{equation}         
where $R_{\rm s}$ is the sphere radius.

\subsection{Cavity radius}\label{CaRad}

The radius of the cavity is found as follows. We approximate the effective collision frequency with
\begin{equation}\label{nueff}
\nu_{\rm eff} = \nu_0\left(\gamma M+\sqrt{\gamma^2M^2+1 }\right),
\end{equation}  
where for argon ions in argon gas $\nu_0\simeq 1.2\times 10^5 P_{\rm Pa}$ ($P_{\rm Pa}$ is the neutral gas pressure expressed in Pa), and $\gamma \simeq 0.23$~\cite{KhrapakJPP2013}. The physics behind Eq.~(\ref{nueff}) is as follows. In a weak electric field, the ion drift velocity is directly proportional to the field, $u\propto E$, and, thus, the effective collision frequency is constant $\nu_{\rm eff}\simeq \nu_0$. In a strong field, however, the drift velocity scales approximately as the square root of the field, $u\propto \sqrt{E}$. This implies $\nu_{\rm eff}\propto \sqrt{E}\propto \nu_0M$. Equation (\ref{nueff}) is constructed to reproduce these two limiting regimes and provides a reasonable interpolation between them using experimental information on drift velocities of Ar$^+$ ions in argon gas (see Appendix~\ref{A0} for a comparison). We substitute this in Eq.~(\ref{ratio1}) and find the critical Mach number $M_*$ corresponding to the condition $|F_{\rm i}/F_{\rm el}|=1$.    
Then using $M=(eE/mv_{\rm {T} i}\nu_{\rm eff})$ together with the long-range asymptote of the electric field (\ref{field}) we finally obtain for the cavity radius
\begin{equation}
R_{\rm cav}\simeq R_{\rm s}\left(\frac{T_e}{T_i}\frac{\ell_i}{R_{\rm s}}\frac{1}{M_*} \right)^{1/2},
\end{equation}
where we have used $\ell_i\nu_{\rm eff}=v_{{\rm T}i}$ and $m_iv_{{\rm T}i}^2=T_i$. The procedure only applies to sufficiently slow drifts, $M_*\lesssim 2$, so that equation (\ref{id4}) for the ion drag force remains adequate.

 {\it A priori} it is difficult to predict correctly the dependence of the cavity size on the neutral gas pressure. If, as one may expect intuitively, $M_*$ is nearly constant (about $M_*\sim 1$), the cavity size should shrink with the increase of the pressure. We shall see, however, in a moment that the actual experimental tendency is opposite and is consistent with the numerical solution of the equations displayed above.      
   
\subsection{Numerical estimates}

For the conditions relevant for the experiments on the injection of milimeter-size metallic spheres in PK-3 Plus facility we adopt the following plasma parameters, based on our previous simulations with the SIGLO-2D code~\cite{ThomasNJP2008,KhrapakPRL2011,KhrapakPRE2012}. The central plasma density depends linearly on pressure and, to a reasonable accuracy, described by $n_0\simeq (1.20+0.11P_{\rm Pa})\times 10^8$, where $n_0$ is in cm$^{-3}$. The electron temperature decreases very weakly with pressure and in the range investigated we can take a fixed value $T_e\simeq 3$ eV. Ions and neutrals are at about room temperature, $T_i\sim T_n\sim 0.03$ eV.   

The force balance model developed is almost independent of the size of the microparticles forming the complex plasma. The only point where the dependence on the particle radius $a$ appears explicitly is when defining the non-linear shielding cloud radius $r_{\rm c}$. Furthermore, this dependence is extremely weak, $r_{\rm c}\propto a^{1/5}$. For this reason we take  a fixed ``average'' radius  $a=3$ $\mu$m, providing a relevant ``logarithmic'' length scale for the mixture of particles present in the experimental chamber (diameter varies from $1.55$ to $14.9$ $\mu$m~\cite{SchwabeNJP2017}).      

\begin{figure}
\includegraphics[width=8cm]{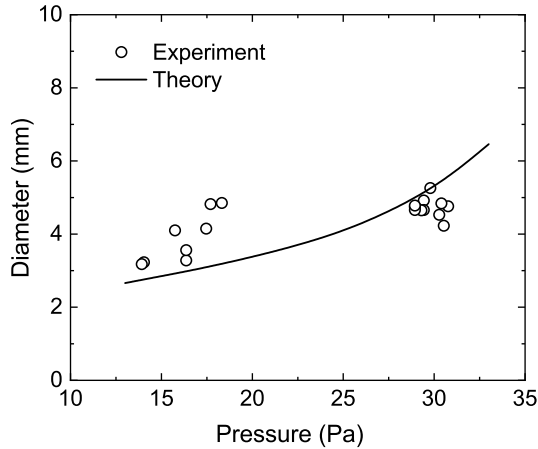}
\caption{Dependence of the cavity diameter on the neutral gas pressure. Circles are experimental measurements (symbol's size is comparable to experimental uncertainty), the solid curve corresponds to the theoretical calculation.}
\label{Fig3}
\end{figure}

With the specified parameters, a numerical calculation is easy to perform. We have first verified that the necessary condition  $|F_{\rm i}/F_{\rm el}|>1$ at $M=0$ is satisfied in the regime investigated. We then estimated $M_*$ and the cavity size as described in Sec.~\ref{CaRad}. The resulting dependence of the cavity diameter on the neutral gas pressure is shown in Fig.~\ref{Fig3}. The agreement with experimental results is reasonable.

Note that on the low-pressure side, the cavity diameter can be underestimated, because the critical velocity from Eq.~(\ref{ratio1}) exceeds $2$ (at $P\lesssim 15$ Pa). This is where the model developed overestimates the ion drag force and hence pushes microparticles closer to the sphere. The actual cavity size can be larger than the theory predicts, as we indeed see in the experiment.   

\section{Discussion}

The experimentally measured cavity size and its dependence on the neutral gas pressure have been demonstrated to be in good agreement with the theoretical approximation developed. It is appropriate to discuss several issues related to limitations and generalizations of the theoretical model.

The cavity size is predicted to be independent (or, at least, very weakly dependent) on the size of microparticles interacting with the big sphere. This is, however, true only when the non-linear model for the ion drag force is appropriate, that is for sufficiently large microparticles when the condition $\beta\gtrsim 1$ is satisfied.

According to the pressure range investigated experimentally, collisional regime for the ions has been considered. Generalization of the approach to the regime of collisionless ions is discussed in Appendix~\ref{A1}.

In the force balance model we neglected the pressure force coming from interparticle repulsion in the complex plasma cloud, because it was previously estimated smaller than the electric and ion drag forces. The pressure force pushes microparticles towards the sphere and, if retained, it would result in somewhat smaller theoretical values for the cavity size.

The specifics of our approach is that we formulate the force balance condition for the weakly perturbed region sufficiently far from the sphere surface. In the original consideration of the force balance, a Yukawa potential around the sphere with the screening length given by the electron Debye radius was assumed inside the cavity~\cite{SchwabeNJP2017}. In addition, the charge of the sphere was required. Our present approximation is considerably simpler in this respect, because the long-range asymptote of the electric field depends only on the electron temperature and sphere radius and is insensitive to other parameters, which are often not known.

The cavity size around a floating object is most sensitive to the electron-to-ion temperature ratio. Since in a typical gas discharge the ion temperature is normally close to the room temperature, while the electron temperature can vary in a relatively wide range, the observation of cavities can potentially be used as a diagnostic tool for the electron temperature.

It should be noted again that the cavity formation is not the only observation from the original experiment. Further interesting phenomena included the formation of bubbles, repulsive attraction (characterization of the long-range ion-drag-mediated attraction of microparticles to the sphere), and wave excitation. These are described in detail in the original paper~\cite{SchwabeNJP2017}.

\section{Conclusion}
  
Interactions between millimeter size floating spheres and a complex plasma have been studied in the PK-3 Plus laboratory onboard ISS. One of the manifestations of these interactions represents the formation of cavities (regions free of microparticles) around the spheres. The cavity size is dictated by the balance of forces acting on the particles at the cavity edge, most important forces being the electric repulsion at short distances and the ion-drag-mediated attraction at long distances. In this article we have proposed a simple practical approach to estimate the ion drag force for experimentally relevant conditions (with the main point to account simultaneously for non-linear ion-particle ineractions and ion-neutral collisions). This has resulted in a simple theoretical approximation for the force balance condition and allowed us to estimate the size of the cavity and its dependence on plasma parameters.
The results of theoretical calculation have been demonstrated to agree well with the experimental results.       
In addition, generalization of the model for the regime of collisionless ions has been made (see below in the Appendix). The theoretical approach reported can be useful in situations when large objects interact with complex plasmas.     
 
\begin{acknowledgments}
We thank Mierk Schwabe and Erich Z\"ahringer for careful reading of the manuscript.  The microgravity research is funded by the space agency of the Deutsches Zentrum f\"ur Luft- und Raumfahrt e.V. (DLR) with funds from the federal ministry for economy and technology according to a resolution of the Deutscher Bundestag under grant No. 50WP0203 and 50WM1203. Support from ROSCOSMOS of the PK-3 Plus project is also acknowledged.
\end{acknowledgments}

 \appendix
 
 \section{Mobility of Ar$^+$ ions in Ar gas}\label{A0}
 
\begin{figure}
\includegraphics[width=7cm]{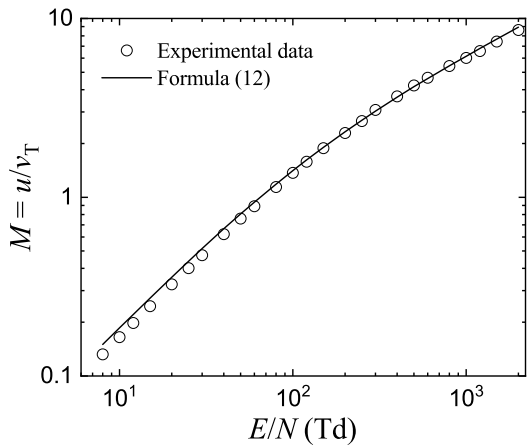}
\caption{Reduced drift velocity of Ar$^+$ ions in Ar gas at $T=300$ K as a function of $E/N$ -- the ratio of electric field strength to the neutral gas density. The latter is measured in Townsend (Td) units; 1 Td = $10^{-17}$ V cm$^2$. Symbols correspond to the experimental data~\cite{Ellis1976}. The curve is calculated using Eqs.~(\ref{drift}) and (\ref{nueff}). }
\label{Fig4}
\end{figure}

Figure~\ref{Fig4} shows the comparison between experimental data on Ar$^+$ ion mobility in Ar gas~\cite{Ellis1976} and the approximation of Eqs.~(\ref{drift}) and (\ref{nueff}). For subthermal ($M<1$) drifts the theoretical approximation slightly overestimate the experimental mobility. For nearly-thermal and superthermal drifts the theory and experiment agree well.
There is clearly some room for improvements, but for the present purposes the accuracy of Eq.~(\ref{nueff}) is quite sufficient.        
 
 \section{Cavity size in the collisionless regime}\label{A1}
 
Let us consider a hypothetical situation of a floating sphere in the collisionless regime for the ion component. This corresponds to the regime $\ell_i\gg R_{\rm s}$, which can be realized at very low pressures. This situation can also be of some relevance and interest in the context of astrophysical plasmas. In this case the long-range asymptote of the electrostatic potential around a sphere is again dictated by the ion absorption on the sphere surface. Quite generally, the potential can be estimated from
\begin{equation}
\phi(r)\simeq - \frac{T_i}{e}\frac{J_0}{J(r)},
\end{equation} 
where $J_0$ is the flux of ions on the sphere surface and $J(r)$ is their influx into the spherical surface of radius $r$~\cite{ChaudhuriSM2011}. (This consideration works also in the collisional case, but in that case we made use of already existing expressions for the potential and electric field~\cite{RaizerBook}). In the case of thin collisionless sheath around a large sphere we have
\begin{equation}
J_0\simeq 4\pi R_{\rm s}^2n_{\rm B}c_{\rm s},
\end{equation}
where $n_{\rm B}\simeq n_0 e^{-1/2}\simeq 0.607n_0$ is the plasma density at the sheath edge and $c_{\rm s}=\sqrt{T_e/m_i}$ is the ion sound velocity. In the weakly perturbed region sufficiently far from the sphere the influx $J(r)$ is simply
\begin{equation}
J(r)\simeq \sqrt{8\pi}r^2n_0v_{{\rm T}i}.
\end{equation} 
This yields 
\begin{equation}\label{pot}
\phi(r)\simeq -1.5\sqrt{\tau}(T_i/e)(R_{\rm s}/r)^2.
\end{equation}
In the case of a smaller object (e.g. microparticle), the orbital motion theory (OML)~\cite{FortovPR2005,Allen1992,Kennedy2003} can be applied to give
\begin{equation}
J_0=\sqrt{8\pi} R_{\rm s}^2 n_0 v_{{\rm T}i}(1+z\tau), 
\end{equation}
and in this regime 
\begin{equation}
\phi(r)=-(T_i/e)(R_{\rm s}/r)^2(1+z\tau)\simeq QR_{\rm s}/r^2.
\end{equation}
The well known long-range $\propto r^{-2}$ asymptote is reproduced~\cite{AlpertBook}. We identify the main difference from the collisional regime: The potential drops faster as $\propto r^{-2}$ instead of $\propto r^{-1}$ decay~\cite{KhrapakPRL2008}. The location of the boundary can be estimated from the condition $M_*\simeq 2$, because in the collisionless regime the ratio $|F_{\rm i}/F_{\rm el}|$ decreases very quickly with $M$. The energy conservation then simply yields
\begin{equation}\label{energy}
-e\phi(r)=\frac{m_i u^2}{2} = 2m_i v_{{\rm T}i}^2 = 2T_i. 
\end{equation} 
Combining (\ref{pot}) and (\ref{energy}) we finally get for the collisionless regime
\begin{equation}
R_{\rm cav}\simeq 0.9R_{\rm s}(T_e/T_i)^{1/4}.
\end{equation} 
This describes the cavity size in the collisionless limit. For $\tau\sim 100$ we arrive at $R_{\rm cav}\simeq 2.8 R_{\rm s}$. This is not very far from the present experimental results on the low-pressure side, see Fig.~\ref{Fig3}.   

\bibliographystyle{aipnum4-1}
\bibliography{Cavity}

\end{document}